\documentclass{ws-ijmpa}
\usepackage[super,compress]{cite}
\usepackage{graphicx}
\usepackage{enumerate}
\usepackage{longtable}
\usepackage{CJK}
\usepackage{lscape}
\usepackage{indentfirst}
\usepackage{float}
\graphicspath{{figures/}}
\begin{document}
\markboth{Authors' Names}{Instructions for typing manuscripts (paper's title)}

%
\catchline{}{}{}{}{}
%

\title{The CEPC Clock Issue and Finetuning of the Circumference\footnote{
supported by Center for High Energy Physics, Henan Academy of Sciences.}
}

\author{Dou Wang\footnote{
corresponding author, wangd93@ihep.ac.cn.}
}

\address{Institute of High Energy Physics, Chinese Academy of Sciences, 19B, Yuquan Road\\
Beijing, 100049, China}

\author{Jie Gao$^{1,2}$, Jianchun Wang$^1$, Dapeng Jin$^1$, Xiaohao Cui$^1$, Ge Lei$^1$, Cai Meng$^1$, Wei Wei$^1$, Jingbo Ye$^1$, Yiwei Wang$^1$, Jiyuan Zhai$^1$, Yuhui Li$^1$, Zhonghui Ma$^1$, Jun Hu$^1$, Shengsen Sun$^1$, Xiaolong Wang$^1$}

\address{1. Institute of High Energy Physics, Chinese Academy of Sciences, 19B Yuquan Road\\
 Beijing, 10049, China\\
2. University of Chinese Academy of Sciences, 19A Yuquan Road\\
 Beijing, 10049, China}

\maketitle
\begin{history}
\received{6th May 2025}
\revised{Day Month Year}
\end{history}

\begin{abstract}
The CEPC clock issue is related with the RF frequency coordination between the various accelerator systems and may affect the operation modes of both the accelerator and the detector. The timing structure of CEPC has been restudied with the collaboration of the accelerator team and the detector team. After discussions between two sides, the CEPC bunch structure is set such that the spacings between adjacent bunches in any CEPC operation mode are integer numbers of 23.08 ns. The master CEPC clock will be provided by the accelerator to the detector systems with a frequency of 43.33 MHz, synchronous to the beam. The CEPC detector system relies on the clock to sample physics signal at the right time. It was found that if the circumference of CEPC is slightly changed to 99955.418 m, not only is the orbit length closer to 100 km, but also the detector would benefit more for the first 10-year operation.

\end{abstract}

\ccode{PACS numbers:29.20.db}


\section{Introduction}	

The Circular Electron Positron Collider (CEPC) is a 100km double ring collider with two interaction points proposed to be working at four energy schemes of Z pole (45.5GeV), W (80GeV), Higgs (120GeV) and tt (180GeV) [1]. The CEPC consists of a linear accelerator (Linac), a damping ring (DR), the Booster, the Collider and several transport lines. The layout of the CEPC accelerator complex is shown in Fig. 1.

A tentative ``10-2-1-5'' operation plan is to run the CEPC first as a Higgs factory for 10 years to produce four million Higgs particles, followed by 2 years of operation as a Super Z factory to create one trillion Z bosons, and then 1 year as a W factory to produce approximately 100 million W bosons. Finally, an upgrade will enable the CEPC to operate at the $t\bar{t}$ energy with 0.6 million $t\bar{t}$  pairs produced. The CEPC's circumference is approximately 100 km, which is compatible with the SppC, a proton-proton collider designed to operate at 125 TeV center-of-mass energy using 20 Tesla superconducting dipole magnets.

A conceptual design has been made for the luminosity goal $3/10×10^{34} cm^{-2} s^{-1}$ for Higgs/W at 30MW and $32×10^{34} cm^{-2} s^{-1}$for Z at 16.5MW in CDR [2]. To mainly increase the luminosity at the Higgs and Z energy, a higher luminosity scheme of the CEPC has been proposed in TDR [1]. The luminosities of the CEPC are mainly limited by the synchrotron radiation (SR) power. The baseline SR power is 30 MW per beam, which is upgradable to 50 MW. For the baseline design with 30 MW SR power, the luminosity at the Higgs mode is $5×10^{34} cm^{-2} s^{-1}$ with 268 bunches, while at the W mode, it is $1.6×10^{35} cm^{-2} s^{-1}$ with 1297 bunches, both with 3T detector solenoid field. At the Z pole, the luminosity is $1.2×10^{36} cm^{-2} s^{-1}$ with 11934 bunches with 2 T detector solenoid field.

\begin{figure}[H]
\centering
 \includegraphics[width=9cm]{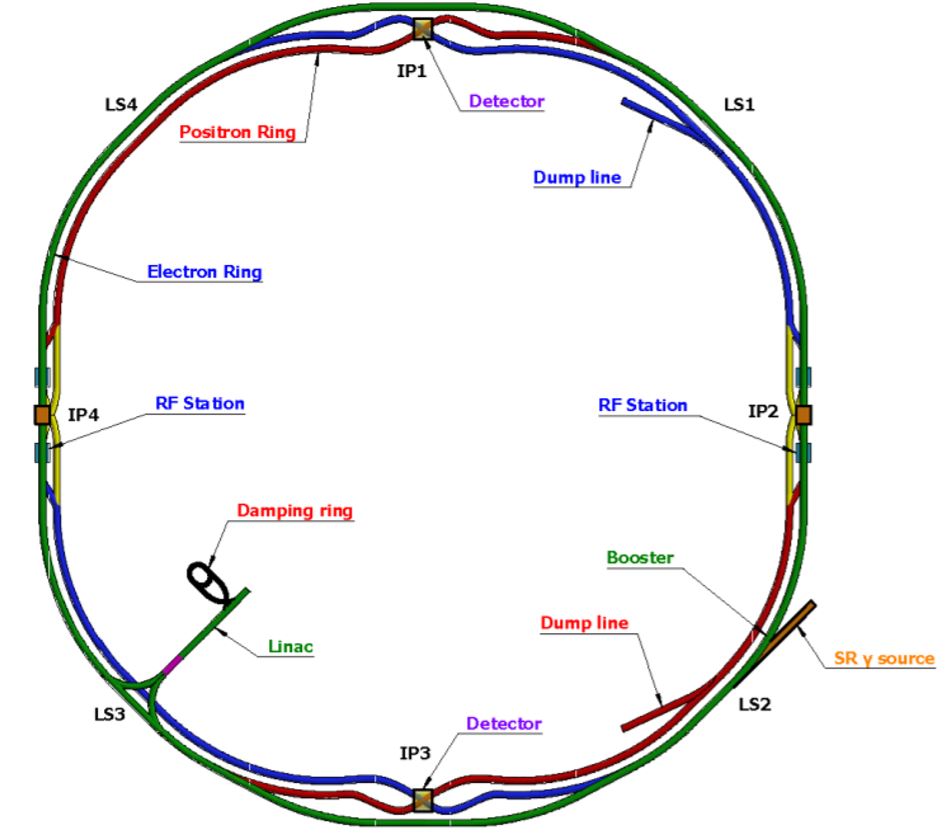}
\caption{Layout of the CEPC accelerator complex.}
\end{figure}

To achieve the designed luminosity, both the top up injection and full injection from empty for the collider ring needs to meet the design requirement. In addition, the hardware of injection and extraction for each subsystem should be compatible with different energy modes. According to the requirements in the collider, the injection chain, injection scheme and the timing structure of CEPC for four energy modes have been studied for a long time [3-8].

Recently, some joint discussions about the overall clock issue and the scheme of synchronized operation between the detector and accelerator has been initiated so that the detector team and accelerator team can collaborate and solve the problems together. It was found that the bunch structure and the circumference of CEPC is related to the feasibility and reliability of the detectors. After careful discussions, a slight adjustment to the circumference was suggested, which is agreed to be beneficial for both the detector and the accelerator.

\section{CEPC accelerator systems and main parameters in TDR }

\subsection{CEPC Main parameters and Collider}

The main parameters of the CEPC [1], as described in the TDR, are listed in Table 1 and Table 2. The luminosities of the CEPC are mainly limited by the synchrotron radiation (SR) power. The baseline SR is 30 MW per beam, which is upgradable to 50 MW. In Table 1, The 10 MW parameters is a transition mode between the Higgs running and formal Z operation using the same hardware at Higgs energy.

\begin{longtable}[htbp]{cc}
  \caption{ CEPC baseline parameters in TDR.}\label{1}\\
  \includegraphics[width=0.92\textwidth]{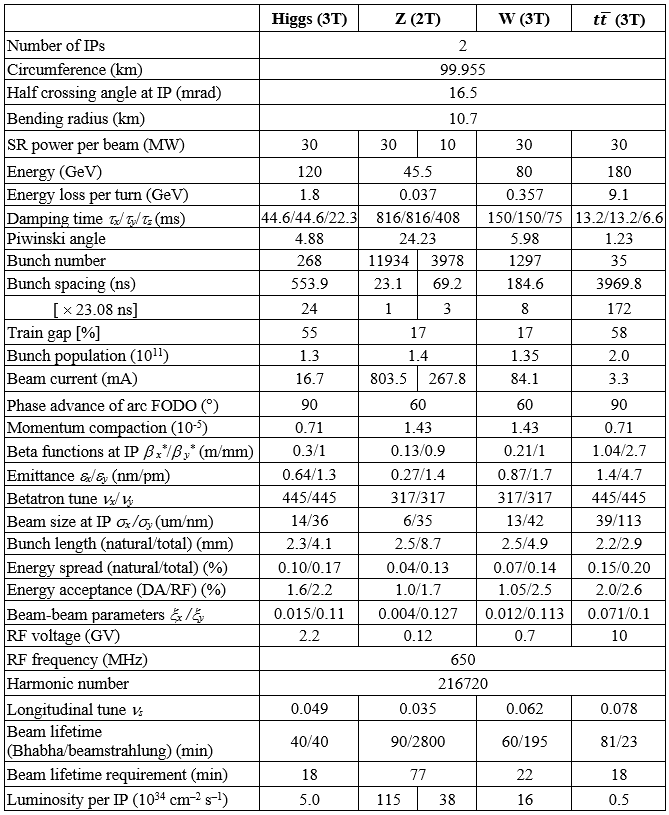}
\end{longtable}

\begin{longtable}[htbp]{cc}
  \caption{ CEPC main parameters with 50 MW upgrade.}\label{1}\\
  \includegraphics[width=1.02\textwidth]{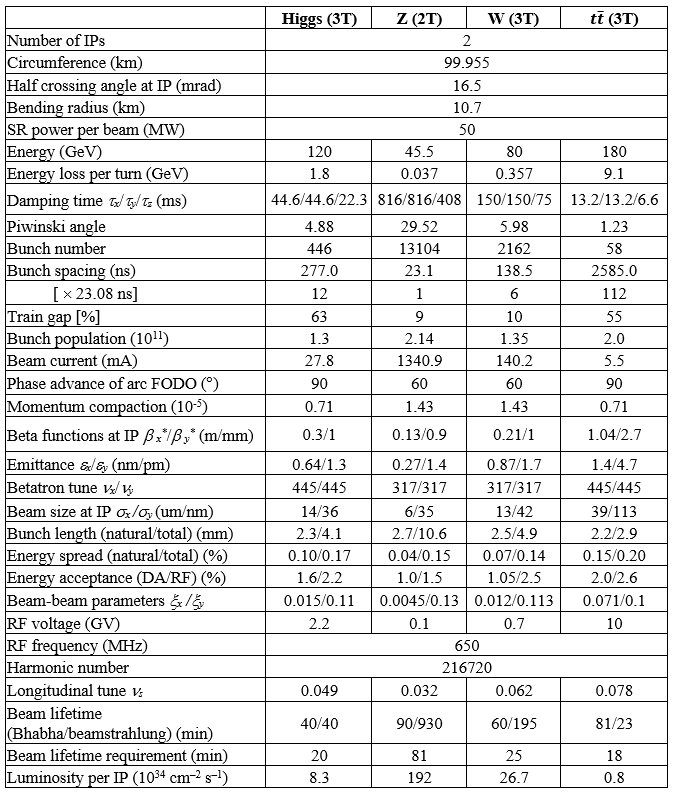}
\end{longtable}

To reach the high luminosity of circular Higgs factory and limiting the construction cost, a partial double ring layout with the crab-waist scheme has been proposed for CEPC in 2013 [9], where electron and positron beams at Higgs and $t\bar{t}$ energies share a common superconducting radio frequency accelerator [10]. While during Z and W energy running, the collider ring is transformed into full double ring layout by bypass beam lines, and electron and positron beam each has its own superconducting radio frequency accelerator. With dedicated design for the bypass beam lines, it is easily switchable between different energy modes without hardware replacement.

In order to improve the luminosity as large as possible, the elaborate crab-waist scheme [11] is adopted with the large Piwinski angle collision [12]. Lattice is carefully optimized in this region and the whole ring with parameters of dipoles, quadrupoles and sextuples accordingly [13, 14]. The machine-detector interface (MDI) issues are one of the most complicate and challenging topics at the CEPC and other future high energy colliders.  The MDI region is about 14m (±7m from the IP) in length in the Interaction Region (IR), where many elements from both detector system and accelerator components need to be installed including the detector solenoid, anti-solenoid, luminosity calorimeter (LumiCal), interaction region beam pipe, cryostat, beam position monitors (BPMs) and bellows. The cryostat includes the final doublet superconducting magnets and anti-solenoid. The CEPC detector consists of a cylindrical drift chamber surrounded by an electromagnetic calorimeter, which is immersed in a $ 2\sim 3 T $ superconducting solenoid of about 9.0 m in length. From the requirement of detector, the conical space with an opening angle should not larger than 8.11 degrees. After optimization, the accelerator components inside the detector without shielding are within a conical space with an opening angle of 6.78 degrees. The crossing angle between electron and positron beams is 33 mrad in horizontal plane. The first final focusing quadrupole is 1.9 m from the IP.

Beamstrahlung is the synchrotron radiation excited by the beam-beam force, which is a significant phenomenon in a storage ring based electron positron collider especially at high energy region [15]. It will increase the energy spread, lengthen the bunch and may reduce the beam lifetime due to the long tail of the photon spectrum. The beamstrahlung lifetime is a major limitation in the $t\bar{t}$ /Higgs mode, so more efforts were dedicated to improving the energy acceptance of the dynamic aperture in the collider lattice design.

The circumference is one of the most important and basic parameters to the CEPC. A comprehensive study has been carried out for the best performance-cost ratio and the most significant prospect [16]. The instant luminosity, construction and operation costs with respect to the circumferences were calculated with a well-established evaluation model. Taking into account of the project overall cost (construction and operation), the objective accumulated particle number, and therefrom the expenditure per particle, it is concluded that the optimal circumference for the Higgs operation is 80 km. Including the highest priority Higgs operation, the Z pole operation, the high-energy upgrade for the top quark factory, and the grand potential project of the SPPC, the circumference of 100 km is the best choice.

\subsection{Linac and Booster}

The Booster is designed to provide electron and positron beams to the Collider at different energies with same circumference as the Collider [1]. The energy range in the Booster is from 30 GeV to the required energy in the Collider. To maintain the required beam current in the Collider, top-up injection is required.  The main Booster parameters at injection and extraction energies are listed in Table 3 and Table 4.

\begin{longtable}[htbp]{cc}
  \caption{ Main Booster parameters at injection energy.}\label{1}\\
  \includegraphics[width=1.02\textwidth]{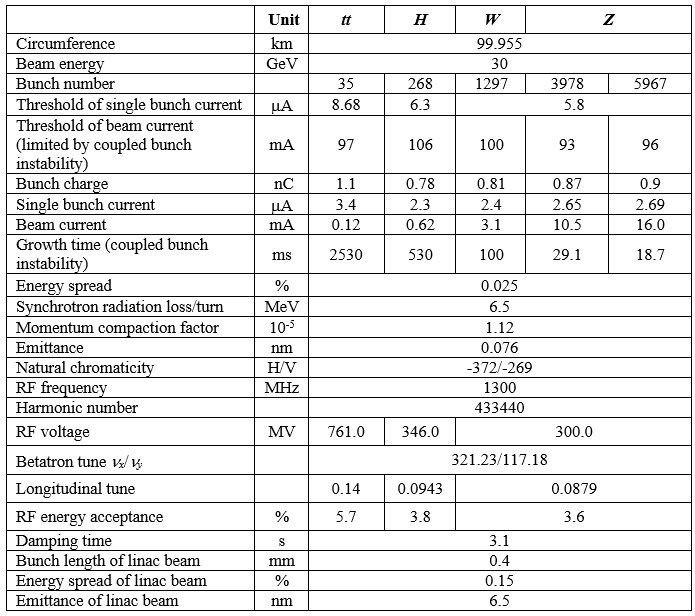}
\end{longtable}

\begin{longtable}[htbp]{cc}
  \caption{Main Booster parameters at extraction energy. }\label{1}\\
  \includegraphics[width=0.92\textwidth]{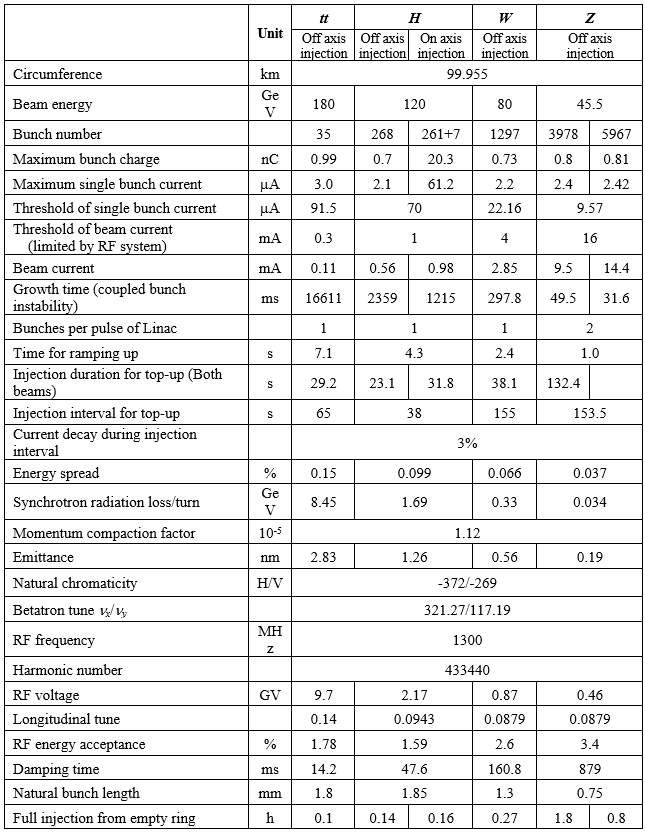}
\end{longtable}

The Booster is injected by a conventional 1.8 km Linac [1]. The Linac system operates at a repetition rate of 100 Hz, and a one-bunch-per-pulse scheme is used. However, for the high luminosity mode at the Z pole (30MW/50MW), a double-bunch scheme is also considered. The CEPC Linac is a type of linear accelerator that uses normal conducting RF technology and operates at two different frequencies, S-band (2860 MHz) and C-band (5720 MHz), as indicated in Table 5. The parameters of S-band accelerating structure and C-band accelerating structure are shown in Table 6. 

\begin{longtable}[htbp]{cc}
  \caption{Main parameters of the Linac.}\label{1}\\
  \includegraphics[width=1.0\textwidth]{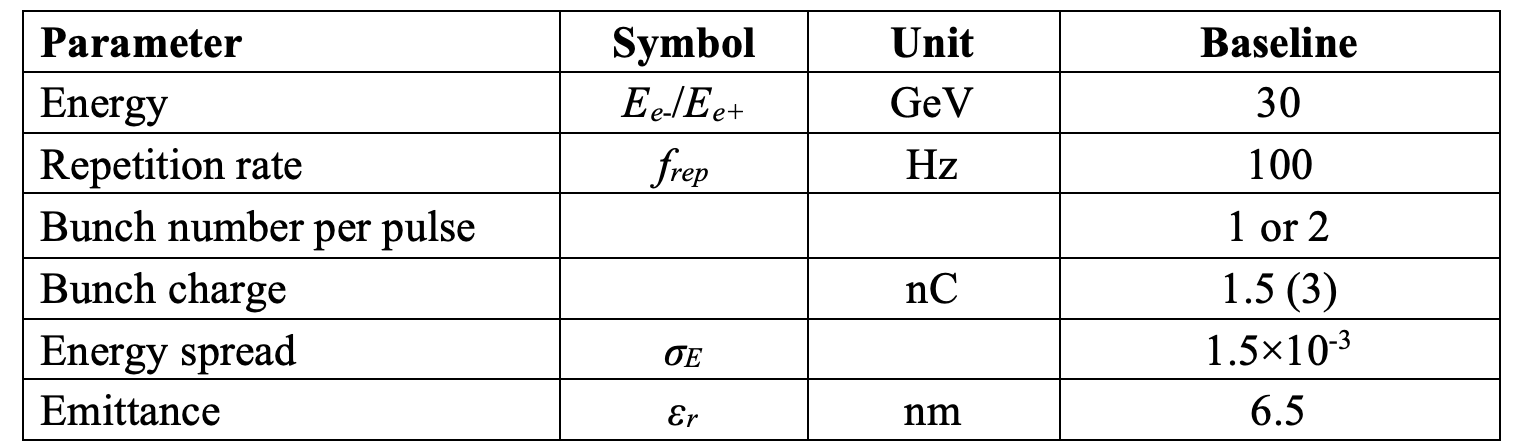}
\end{longtable}

\begin{longtable}[htbp]{cc}
  \caption{ Main Parameters of the Linac Accelerating Structures.}\label{1}\\
  \includegraphics[width=1.0\textwidth]{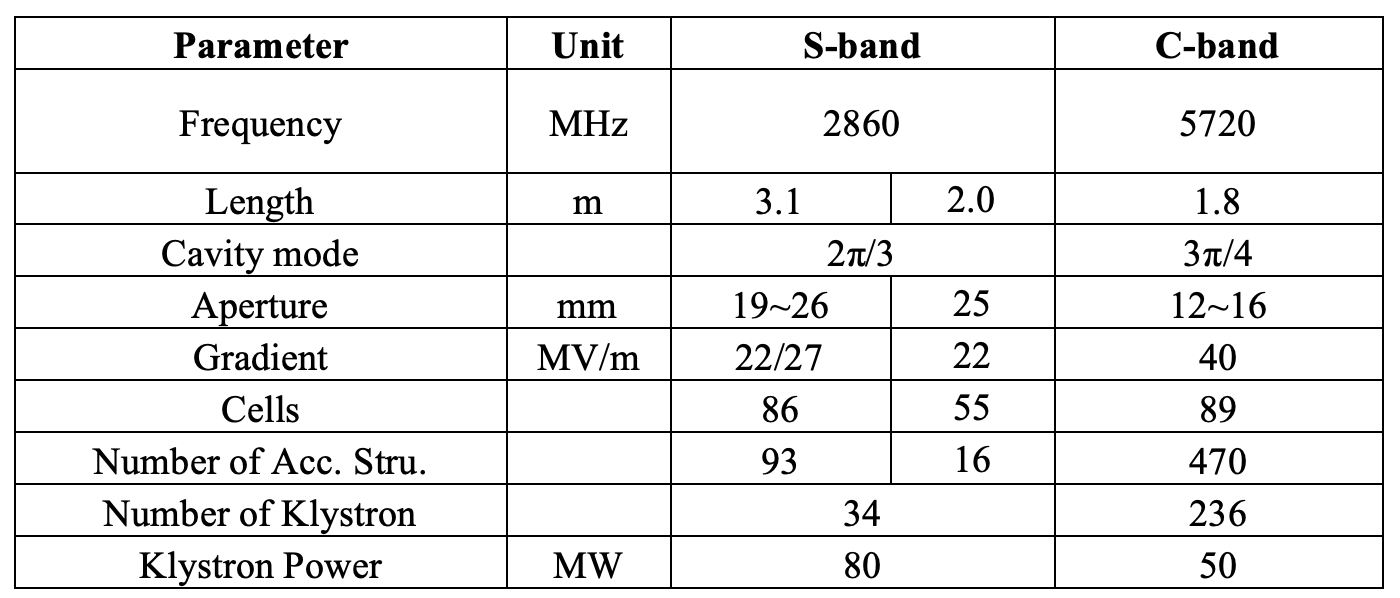}
\end{longtable}

The Linac system also includes a positron Damping Ring (DR) at 1.1 GeV with 147.5 m circumference. The main purpose of the DR is to reduce the transverse emittance of the positron beam at the end of the Linac. The RF frequency of the DR is 650MHz which is same as the Collider. The main parameters of the damping ring are listed in Table 7.

\begin{longtable}[htbp]{cc}
  \caption{Main Parameters of the Damping Ring.}\label{1}\\
  \includegraphics[width=1.0\textwidth]{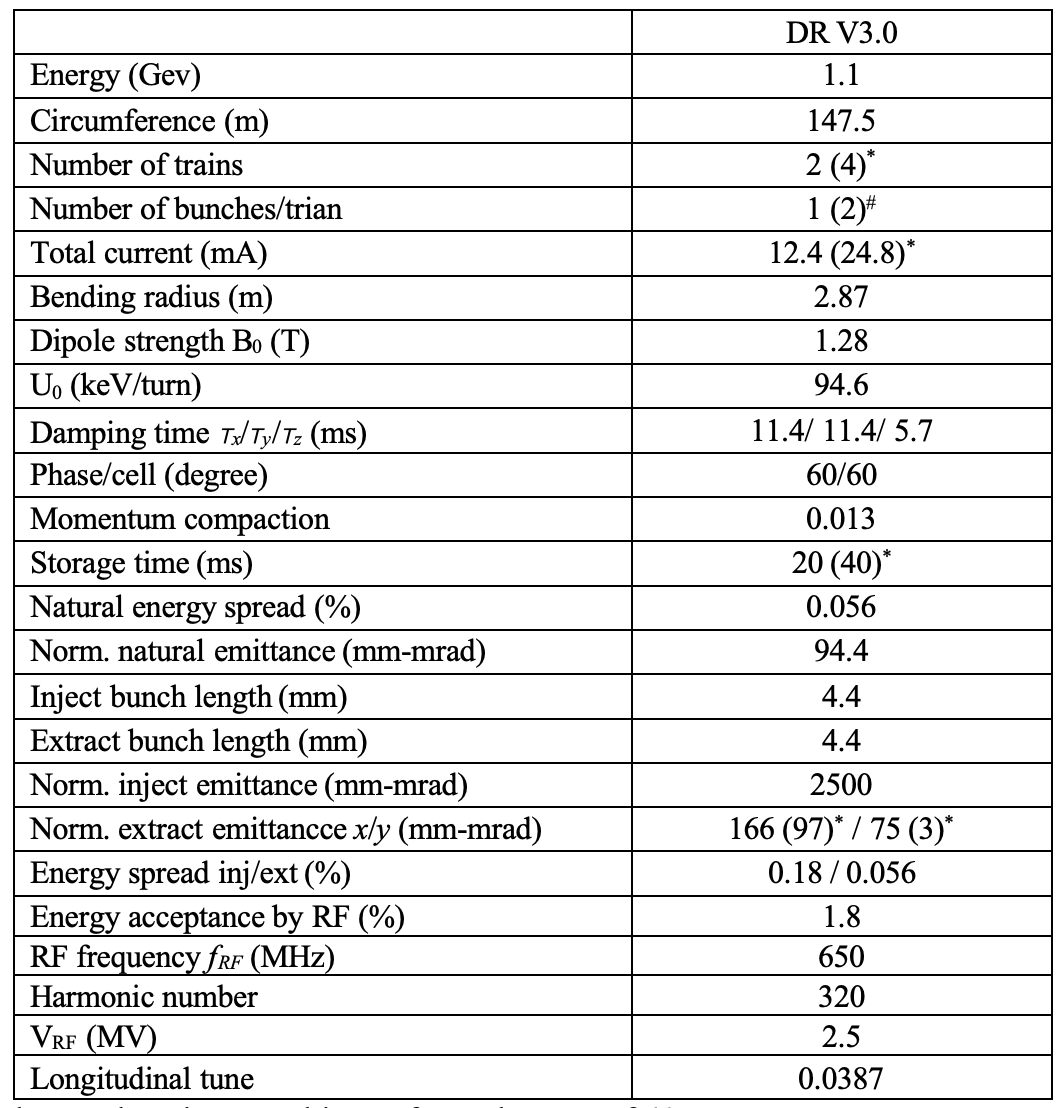}
\end{longtable}

$*$  The numbers in parenthises refer to the case of 40 ms storage.

$\#$ The number in the parentheses is for the Linac double-bunch operation at Z pole.

\section{CEPC frequency choice in TDR }	

The RF frequency choice in TDR is shown in Table 8. SHB1 and SHB2, in this table, denote the sub-harmonic buncher after the electron source. The RF frequencies for Linac, Booster, and Collider Rings are 2860MHz for S-band accelerating structures, 5720MHz for C-band accelerating structures, 1300MHz and 650 MHz, respectively. The GCD (Greatest Common Divisor) is 130 MHz. Thus the minimum bunch spacing is 7.692 ns derived from a 130 MHz clock.

\begin{longtable}[htbp]{cc}
  \caption{CEPC RF frequency parameters in TDR.}\label{1}\\
  \includegraphics[width=1.0\textwidth]{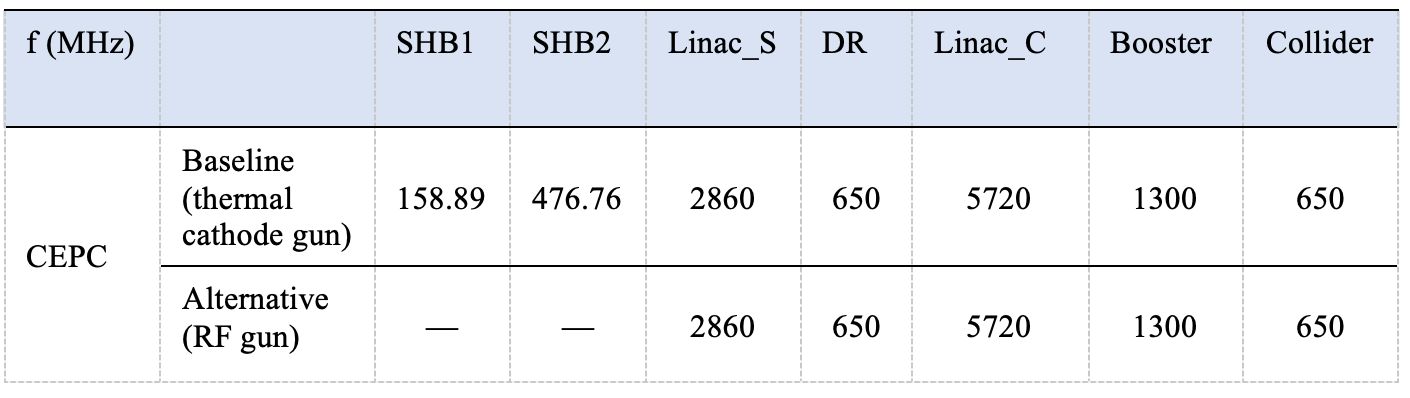}
\end{longtable}

\section{ CEPC minimum bunch spacing and detector frequency choice.}

The minimum bunch spacing for CEPC is related to both accelerator requirement and detector requirement. For CEPC, Z pole is the most demanding mode with the largest bunch number to reach the design luminosity. The bunch separation for Z mode should be shorter than about 25.4ns in order to accommodate enough bunches, which is shown in Equ. (1), where C is the circumference, N is the bunch number for 50 MW Z mode. On the other hand, the bunch separation should not be too small from the detector side because the frequency of detector should not be too high, otherwise the power consumption, data transmission, and spill-over effect in the sub-detector system can be problematic. In addition, there is also a truth that most electronic components and detector technologies developed at CERN aiming to serve the LHC experiments are optimized to a base frequency of 40 MHz. To avoid re-invention or re-optimization and allow maximal use of “commonly” available components, CEPC detector prefer to use a similar base frequency, providing that the impact to the performance is under control. After careful discussion between accelerator side and detector side, it is agreed to have the detector base frequency of 43.3 MHz which is one third of 130 MHz.

\begin{equation}
    T_{max}=\frac{C}{cN}=\frac{100km}{c*13104}=25.4ns
\end{equation}

CEPC has four energy modes (H, Z, WW, and $t\bar{t}$ ) and the frequencies of the clock for the 4 different operation modes need to be the same so that all detectors and electronics can be optimized around one frequency in design. Thus the minimum bunch spacing for Z pole is 23.08 ns according to the detector frequency of 43.3 MHz. And the bunch spacing of all the other modes need to be an integer of 23.08 ns. Concerning the synchronization issue, a 43.3 MHz clock should be provided by the accelerator to the experiments, synchronized with the bunches. The detectors will lock to this beam clock with PLL/DLL and sample signal at pre-fixed optimal phases.

\section{CEPC circumference related to the clock issue}

Since all detector activities should be synchronized by the same clock to sample the physics signal at the right time. Not only the bunch spacing of all the energy modes should be an integer of 23.08 ns, but also the revolution time of CEPC should be an integer of 23.08 ns.

Accelerator recommended a group of setting for bunch structure based on the requirement for luminosity and bunch number with 23.08 ns as the minimum bunch spacing which are listed in table 1 and table 2. Table 1 listed the CEPC main parameters in TDR with 30 MW SR power per beam. The 10 MW case is added in table 1 to consider the transition mode between the 10-year Higgs running and following Z operation. The advantage of the transition 10 MW Z mode is that both accelerator and detector can use the same hardware at Higgs energy without any upgrade and system tuning, so that it is easily switchable between Higgs and Z modes. The bunch structure in table 9 and table 10 can fulfill the timing requirement of both accelerator and detector.

According to the bunch structure in table 9 and table 10, the circumference of CEPC can be calculated by the following formula
\begin{equation}
    14442\times 23.08ns\times c=99913.908m
\end{equation}
where c is the speed of light.

\begin{longtable}[htbp]{cc}
  \caption{CEPC bunch structure for TDR baseline.}\label{1}\\
  \includegraphics[width=1.0\textwidth]{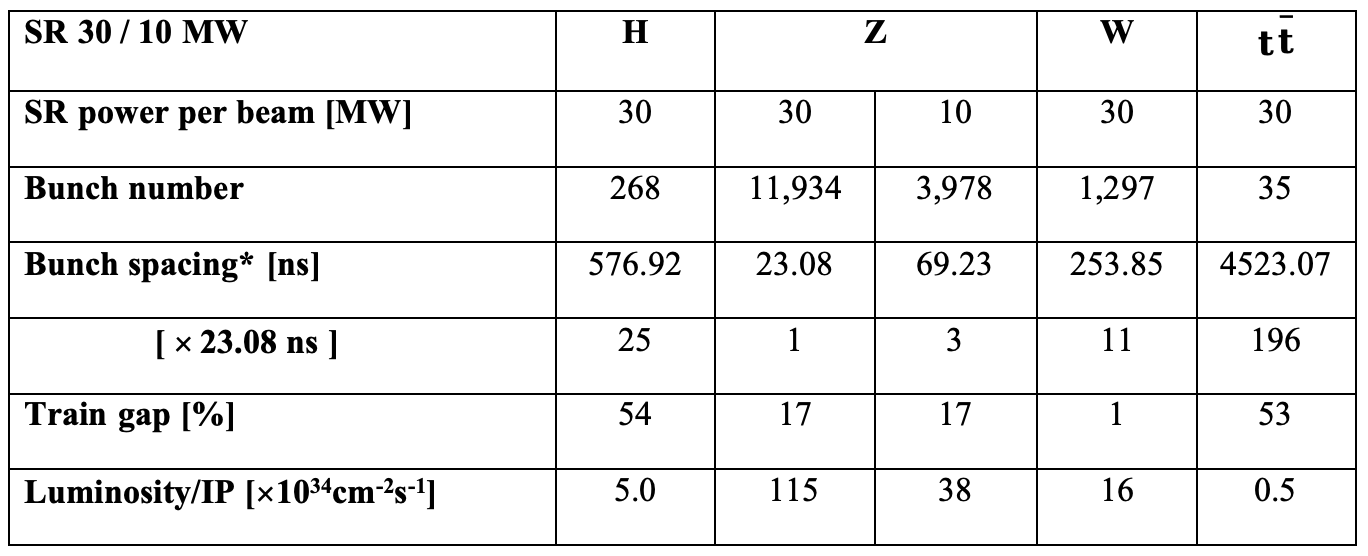}
\end{longtable}

\begin{longtable}[htbp]{cc}
  \caption{CEPC bunch structure for 50 MW upgrade.}\label{1}\\
  \includegraphics[width=1.0\textwidth]{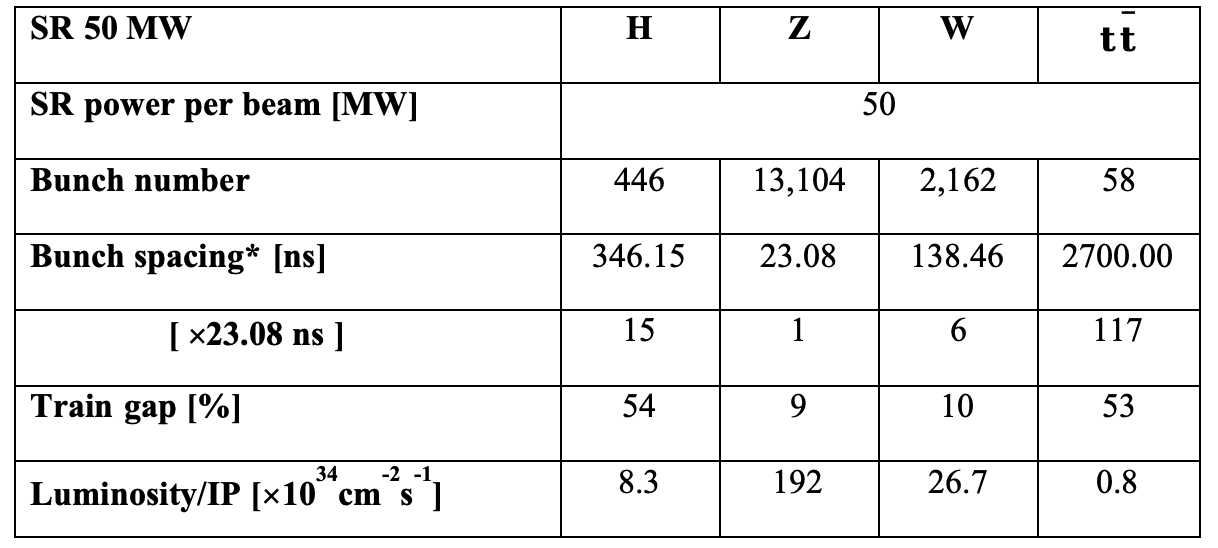}
\end{longtable}

\section{Circumference finetuning and detector performance improvement}

CEPC will operate in the Higgs and the low luminosity (LL) Z modes in the first 10 years. The bunch spacing for the LL Z mode is 3×23.08 ns. If the bunching spacing for the Higgs mode is an integer number of 3×23.08 ns, e.g. 24×23.08 ns, the detector system could benefit more.

The operation mode of the CEPC accelerator has a critical impact on the design and performance optimization of the detector electronics. Two cases of power estimation for the CEPC Vertex detector CMOS sensor are compared:

\begin{itemize}
    \item Case 1: The average interval between accelerator bunches is 69.23ns, and there is a possibility of collision between adjacent bunches with a 23.08ns interval.

    To distinguish collision events between adjacent bunches, the response speed of the detector and the shaping time of the electronics need to be optimized according to the shortest bunch interval of 23.08ns, typically resulting in higher power consumption.

    \item Case 2: The accelerator bunches have a fixed interval of 69.23ns, with no possibility of smaller interval collisions.  

    In this case, the shortest bunch interval becomes 69.23ns, allowing the highest response speed of the detector and electronics to be reduced to one-third, thereby saving a significant amount of power.
\end{itemize}

Due to limitations in cooling space, the vertex detector is most sensitive to power consumption value. As an example, two different power consumption estimates are provided for the two modes in table 11.

\begin{longtable}[htbp]{cc}
  \caption{ Power estimation of the CEPC Vertex Detector CMOS Sensor.}\label{1}\\
  \includegraphics[width=1.0\textwidth]{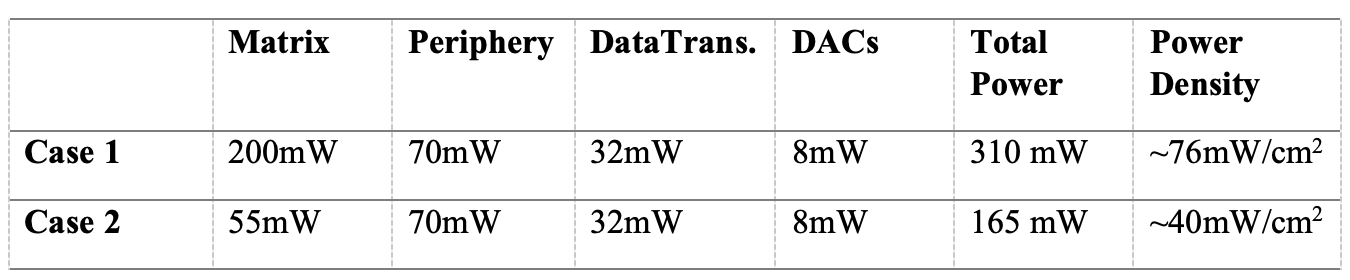}
\end{longtable}

In table 11, the Matrix section mainly represents the chip pixel array, which is the main analog part and the primary item for power optimization. For Case 2, power consumption can be reduced to one-fourth of Case 1.

The remaining items in the table are related to the average counting rate of the detector. Since the average interval between bunches in Case 1 and Case 2 are both 69.23ns, maintaining the same counting rate level (excluding the differences introduced by the MDI background in the two modes), the power consumption of these parts remains unchanged. The output data rate of the chip in both cases is 1Gbps on average.

Considering all factors, the average power consumption of the chip in the two modes is approximately 76mW/cm$^2$ and 40mW/cm$^2$, respectively. These two power levels will also bring significant differences to the cooling and mechanical design of the vertex detector. 

In a word, the performance optimization brought by Case 2 will increase the overall feasibility and reliability of the detector. After discussion between the detector team and accelerator team, it was found that if the number of bunch positions is 14,448 instead of 14,442, not only is the orbit length closer to 100 km, but also the detector would benefit more for the first 10-year operation, i.e. Higgs and LL Z modes. Accordingly, the CEPC bunch structures are modified slightly as table 12 and table 13. With the new bunching structure, the overall power consumption of the Vertex Detector can be reduced to ~53\%. This will significantly reduce the difficulty of cooling scheme and the material budget. It will also benefit the determination of event start time.

\begin{longtable}[htbp]{cc}
  \caption{ CEPC bunch structure for 30/10 MW operation with circumference finetuning.}\label{1}\\
  \includegraphics[width=1.0\textwidth]{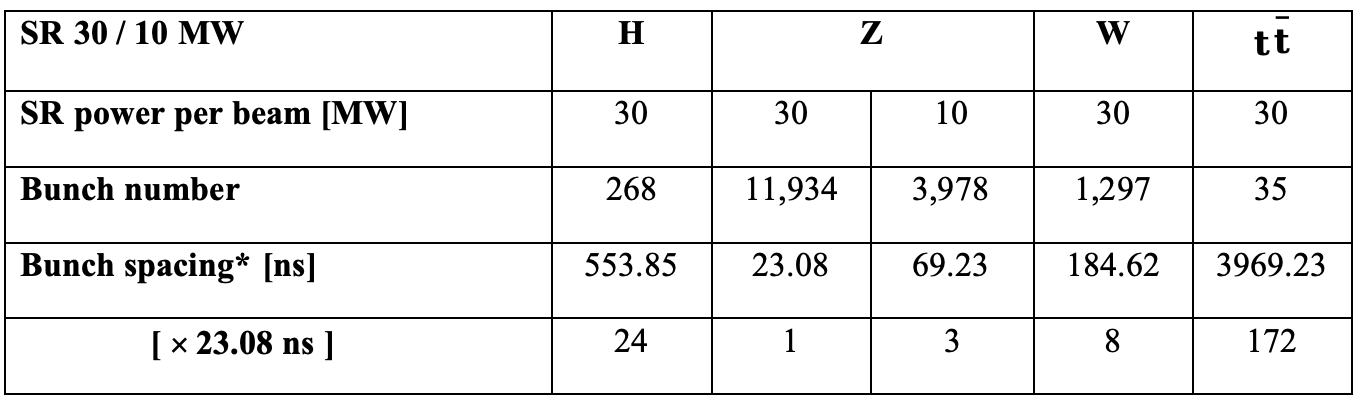}
\end{longtable}

\begin{longtable}[htbp]{cc}
  \caption{ CEPC bunch structure for 50 MW operation with circumference finetuning.}\label{1}\\
  \includegraphics[width=1.0\textwidth]{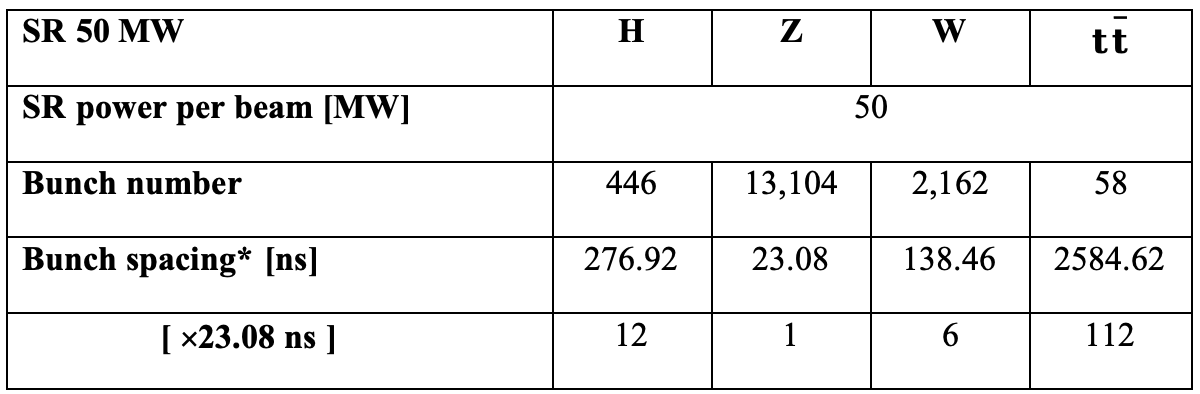}
\end{longtable}

With the modified bunch structures, the circumference of CEPC can be calculated by the following formula
\begin{equation}
    14448\times 23.08ns\times c=99955.418m
\end{equation}
where c is the speed of light.

\section{Summary}

The RF frequencies for Linac, Booster, and Collider Rings are 2860 MHz (and 5720 MHz), 1300 MHz and 650 MHz, respectively. The CEPC clock issue is related with the RF frequency coordination between the various accelerator systems and may affect the operation modes of both the accelerator and the detector. The timing structure of CEPC has been restudied with the collaboration of the accelerator team and the detector team. After discussions between two sides, the CEPC bunch structure is set such that the spacings between adjacent bunches in any CEPC operation mode are integer numbers of 23.08 ns. The master CEPC clock will be provided by the accelerator to the detector system(s) with a frequency of 43.33 MHz, synchronous to the beam. The CEPC detector system relies on the clock to sample physics signal at the right time. To achieve this, the orbit length is adjusted to house 14,442 bunch positions which is 99913.908 m. Furthermore, it was found that if the circumference of CEPC is slightly changed to 99955.418, not only is the orbit length closer to 100 km, but also the detector would benefit more for the first 10-year operation.

\section*{ACKNOWLEDGMENTS}
The authors of this paper thank the great efforts of both CEPC accelerator team and CEPC detector team.


\begin{thebibliography}{99}


\bibitem{1}Gao, J. CEPC Technical Design Report: Accelerator. Radiat Detect Technol Methods (2024). https://doi.org/10.1007/s41605-024-00463-y.            


\bibitem{2}The CEPC-SPPC Study Group, \textit{CEPC Conceptual Design Report,Volume I- Accelerator}. IHEP-CEPC-DR-2018-01, IHEP-AC-2018-01, August 2018..

\bibitem{3}Dou Wang et. al., Problems and considerations about the injection philosophy and timing structure for CEPC, International Journal of Modern Physics A, Vol. 37, (2022) 2246006.

\bibitem{4}Dou Wang, Problems and considerations about the timing \& bunch pattern, CEPC day, Aug. 2021.


\bibitem{5}Cai Meng, Timing and bunch pattern compatibility study for multi RF frequencies, CEPC day, Jan. 2022.            


\bibitem{6}Cai Meng, Preliminary discussion on the timing and injection sequence of the CEPC, internal note, Apr. 2022.

\bibitem{7}Cai Meng, CEPC Linac injector design status, CEPC2022 Workshop, Oct. 2022.

\bibitem{8}Xiaohao Cui, CEPC Transport lines and Timing, CEPC2023 Workshop, Oct. 2023.


\bibitem{9}J. Gao, ultra-low beta and crossing angle scheme in CEPC lattice design for high luminosity and low power, internal note, IHEP-AC-LC-Note2013-012, Jun. 2013           


\bibitem{10}K. Oide et. al., Design of beam optics for the future circular collider e+e- collider rings, Physical Review Accelerators and Beams, vol. 19, no. 11, p. 111005, Nov. 2016. doi:10.1103/PhysRevAccelBeams.19.111005.

\bibitem{11}P. Raimondi, D. Shatilov, M. Zobov, beam-beam issues for colliding schemes with large Piwinski angle and crabbed waist, LNF-07/003 (IR), 29 Jan. 2007.

\bibitem{12}D. Wang et. al., CEPC partial double ring scheme and crab-waist parameters, International Journal of Modern Physics A, vol. 3, p. 1644016, 2016.


\bibitem{13}Y.W. Wang et. al., Lattice design for the CEPC double ring scheme, International Journal of Modern Physics A, Vol. 33, No. 2, (2018)1840001.          


\bibitem{14}Y.W. Wang et al, Lattice Design of The CEPC Collider Ring for a High Luminosity Scheme, WEPAB033, IPAC21.

\bibitem{15}V. I. Telnov, Restriction on the energy and luminosity of e+e- storage rings due to beamstrahlung, Phys. Rev. Letters 110,114801(2013).

\bibitem{16}Dou Wang, et al., CEPC cost model study and circumference optimization, Journal of Instrumentation, 17 P10018 2022.

\end{thebibliography}
\end{document}